# Title: Charting a finite element, mechanical atlas of dermatologic wound closure

Author: Congzhou M Sha[1,2,*]

[1]Department of Engineering Science and Mechanics, Penn State University, University Park, PA, USA

[2]Department of Pharmacology, Penn State College of Medicine, Hershey, PA, USA

[*]Corresponding author: cms6712@psu.edu, ORCID 0000-0001-5301-9459

**Word count: 2,888 (Introduction-Discussion)**

**ABSTRACT (135 WORDS)**

Wound geometry and the mechanical properties of human skin govern the failure modes of partially healed or scarred tissue. Though dermatologists and surgeons develop an intuitive understanding of the mechanical characteristics of skin through clinical practice, finite element models of wounds can aid in formalizing intuition. In this work, we explore the effect of wound geometry and primary intention closure on the propagation of mechanical stresses through skin. We use a two-layer, orthotropic, hyperelastic model of the epidermis, dermis, and subcutis to accurately capture the mechanical and geometric effects at work. We highlight the key assumptions which must be made when modeling closure of wounds by primary intention, clearly delineating promising areas for model improvement. Models are implemented in DOLFINx, an open-source finite element framework, and reference code is provided for reproducible and extensible science.

## INTRODUCTION

The skin is a common site of mechanical trauma. Additionally, in surgical procedures, the skin is physically disrupted to gain access to vasculature, muscles, organs, or other parts of the body. For skin to regain its mechanical integrity with minimal scarring, the wound edges are approximated under tension by sutures, staples, adhesives, or bandages; this is known as healing by primary intention[1]. It is important for clinicians to understand the effect of potential mechanical stresses which may be applied to the wound, to minimize the risk of wound dehiscence, tissue strangulation, and other complications of wound closure.

The finite element method (FEM) is a framework to investigate continuum mechanics and partial differential equations[2]. Because it is difficult and harmful to patients and animals to perform mechanical stress-testing of wounds at different stages of healing, FEM can fill in the gaps in our knowledge. The goal is to create a computational framework for answering complicated clinical questions: "If my patient has a 2 cm elliptical wound on her upper back, with the major axis at a 25-degree angle to a Langer line, where should I concentrate stitches so that when my patient itches her back, she won't reopen the wound?"

In this work, we use the FEM[3] to investigate the effect of wound geometry on the stress during primary intention (Figure 1). We treat the skin as two layers, focusing on small patches of skin, to avoid the burden of modeling of other tissues, but reproducing the key mechanical features of wounds. We focus on mechanical equilibrium as the crucial starting point for further modeling. We improve upon previous work[4] by incorporating 3D modelling, contact mechanics, and performing extensive benchmarking with 68,806 simulations.

**MATERIALS AND METHODS**

*Creating wound geometries*

We used the Python interface to Gmsh 4.11[5] to create and mesh wound geometries. For the geometry, we used rectangular cuboids, whose external faces meet at right angles, except for the wound surface. We used the right-handed 3D Cartesian coordinate system, with the layers of the skin parallel to the $xy$-plane, and $z$ coordinates going from more negative to more positive representing going from deep to superficial. The lateral external faces were parallel to the $xz$- or $yz$-planes. We used first-order Lagrange elements for all calculations. The parameters we used for the geometries are included in Table 1.

*Nonlinear solving*

We used Newton's method with adaptive relaxation for nonlinear solving. To find a root of any vector $F$ of piecewise differentiable functions, where $F_k = f_k(x)$, Newton's method iteratively improves upon an initial guess $x_0$:

$$x_{i+1} = x_i - \gamma \cdot J(x_i)^{-1} F(x_i),$$

(1)

where $J_{ki} = \frac{\partial f_k}{\partial x_i}$ is the Jacobian matrix of $F$, and $\gamma$ is the relaxation parameter. In practice, one defines $\Delta x_i = x_{i+1} - x_i$ and solves

$$J(x_i)\Delta x_i = -F(x_i)$$

(2)

for $\Delta x_i$ to avoid computing matrix inverses[2]. Instead of choosing $\gamma$ arbitrarily, we follow the fast inertial relaxation engine[6], allowing $\gamma$ to increase when the updates were too small and decreasing when the updates were too large. All Dirichlet BCs were strongly enforced.

*Variational formulation of hyperelasticity*

We sought to find mechanical equilibrium for the wound geometries[2]. Skin is mostly water and thus nearly incompressible; additionally, the deformations that occur when skin is approximated by primary intention are large and thus the underlying geometric problem is inherently nonlinear. Therefore, we used an orthotropic, hyperelastic constitutive model for the skin[7].

Consider a wound and its surrounding skin as a subdomain $\Omega \in \mathbb{R}^3$, with each point within $\Omega$ represented by a position $x$. We wish to find the deformation $x \to x + u(x)$ which minimizes a particular hyperelastic strain energy density $W(u(x))$. In terms of the undeformed coordinates $x$, the deformation gradient $\boldsymbol{F}$ is defined as:

$$\boldsymbol{F} = \boldsymbol{I} + \nabla u,$$

(3)

with $\boldsymbol{I}$ the identity matrix and $(\nabla u)_{ij} = \partial_j u_i$. The strain energy gives rise to the constitutive equation

$$\boldsymbol{P} = \frac{\partial W}{\partial \boldsymbol{F}},$$

(4)

where $\boldsymbol{P}$ is the first Piola-Kirchhoff stress tensor, and the force balance within $\Omega$ (no body forces) is

$$\nabla \cdot \boldsymbol{P} = 0.$$

(5)

The divergence of a matrix is defined on each column ($\nabla \cdot \boldsymbol{P} = \sum_j \partial_j \boldsymbol{P}_{ij} = \partial_j \boldsymbol{P}_{ij}$, using the Einstein summation convention).

In the weak formulation of Eq. ( 5 ), we require that for any piecewise differentiable vector function $\eta$,

$$\int_\Omega (\nabla \cdot \boldsymbol{P}) \cdot \eta \, \mathrm{dV} = 0.$$

(6)

Integrating Eq. (6) by parts:

$$\int_\Omega (\nabla \cdot \boldsymbol{P}) \cdot \eta \, \mathrm{dV} = -\int_\Omega \boldsymbol{P} : \nabla \eta \, \mathrm{dV} + \int_{\partial\Omega} (\boldsymbol{P} \cdot \hat{n}) \cdot \eta \, \mathrm{dS} = 0,$$

(7)

where dV is the volume element, $\partial\Omega$ is the boundary of $\Omega$, $\hat{n}$ is the outward facing normal of the boundary, dS is the surface element, and the colon denotes the inner product between tensors ($a:b = \sum_i \sum_j a_{ij} b_{ij}$). The term $\boldsymbol{P} \cdot \hat{n}$ is redefined as the traction $T$, which is specified as a natural boundary condition (BC). We rewrite Eq. (7) as:

$$\int_\Omega \boldsymbol{P} : \nabla \eta \, \mathrm{dV} - \int_{\partial\Omega} T \cdot \eta \, \mathrm{dS} = 0.$$

(8)

The constitutive variational form is:

$$F_{\text{constitutive}} = \int_\Omega \boldsymbol{P} : \nabla \eta \, \mathrm{dV}.$$

(9)

Assuming Eq. ( 8 ) is stationary for all $\eta$ is equivalent to the Euler-Lagrange equations in Eq. ( 5 )[8]. Upon discretization, integrals are converted to sums, and the goal is to minimize the left-hand side, using the automatic differentiation framework of DOLFINx[2].

*Constitutive model of the skin*

> Assumption 1: We used an isotropic, hyperelastic (Mooney-Rivlin) model as our constitutive equation for $W(u(x))$ in the subcutis, and an orthotropic, hyperelastic (Fung) model for the dermis and epidermis, as proposed previously by Flynn et al[7] in modeling of the soft tissues of the face.

The Mooney-Rivlin energy is defined as:

$$W_{MR} = C_{10}(\tilde{I}_1 - 3) + C_{20}(\tilde{I}_1 - 3)^2 + \frac{1}{2}K(\ln J)^2,$$

$$\tilde{I}_1 = \operatorname{tr} \widetilde{\boldsymbol{B}},$$

$$\widetilde{\boldsymbol{B}} = \widetilde{\boldsymbol{F}}\widetilde{\boldsymbol{F}}^T,$$

$$\widetilde{\boldsymbol{F}} = J^{-1/3}\,\boldsymbol{F},$$

$$J = \det \boldsymbol{F},$$

( 10 )

with $F$ the deformation gradient from Eq. (3), $J$ its Jacobian determinant, $\widetilde{F}$ is the distortional part of $F$, $\widetilde{B}$ is the distortional part of the left Cauchy-Green tensor, $\tilde{I}_1$ is the first invariant of $\widetilde{B}$, $C_{10}$ and $C_{20}$ are empirical coefficients, and $K$ is the bulk modulus.

The Fung energy is defined as:

$$W_{Fung} = \frac{1}{2}c(e^Q - 1) + \frac{K}{2}(\ln J)^2,$$

$$Q = \frac{1}{c}\sum_{a=1}^{3}\left[2\mu_a A_a^0 : \widetilde{E}^2 + \sum_{b=1}^{3} \lambda_{ab}(A_a^0 : \widetilde{E})(A_b^0 : \widetilde{E})\right],$$

$$\widetilde{E} = \frac{1}{2}(\widetilde{F}^T\widetilde{F} - I),$$

$$A_a^0 = a_a^0 \otimes a_a^0$$

(11)

where $a_a^0$ is a unit material axis vector, $A_a^0$ is the matrix with respect to which $\widetilde{E}$ is contracted, $\widetilde{F}$ is the distortional part of the deformation gradient, $I$ is the $3 \times 3$ identity matrix, $\widetilde{E}$ is the distortional part of Green-Lagrange strain, and $c, \mu_a, \lambda_{ab}$ are material-specific coefficients. In the Fung model, $a_1^0$ is the direction of a Langer line, $a_3^0$ is the skin surface normal, and the third axis is determined by the cross product of the first two $a_2^0 = a_1^0 \times a_3^0$.

We differentiate $W_{MR}$ and $W_{Fung}$ with respect to the components of the deformation gradient to form the first Piola-Kirchhoff stress tensors $\boldsymbol{P}_{MR}$ and $\boldsymbol{P}_{Fung}$ for Eq. (8). In practice, the DOLFINx framework performs automatic differentiation so that analytic expressions are unnecessary[9–11].

*Approximating* in vivo *skin tension using the appropriate boundary conditions*

<u>Assumption 2: To account for the effect of intrinsic skin tension, Flynn et al defined a scaling of the coordinates of the deepest layer of their models, called the tension scaling factor (TSF)[7]. In our models, we applied the same tension scaling factor.</u>

We enforced the TSF using Dirichlet BCs on the deformation $u$. An external face described by $x = a$ is parallel to the $yz$-plane, and the BC is:

$$u_x = \text{TSF} \cdot a.$$

(12)

Similarly, the external faces $y = b$ receive a BC of:

$$u_y = \text{TSF} \cdot b.$$

(13)

To anchor the skin in the $z$ direction and eliminate the translational symmetry of the system, we set a BC of:

$$u_z = 0$$

(14)

on the deep face of the skin. We will refer to the boundary conditions in Eqs. (12) and (13) as $DBC_{\text{lateral tension}}$, and Eq. (14) as $DBC_{\text{bottom}}$. We used the hyperelastic coefficients and TSFs from Flynn et al[7], as listed in Table 2.

The tension scaling can be replaced by a traction $T$ on the lateral faces, which can be obtained by first solving Eq. (8) for $u$, calculating $P$, and then taking the inner product of $P$ with the surface normal

$$T = P \cdot \left(J F^{-T} \widehat{N}\right).$$

(15)

$JF^{-T}$ reduces to the identity here because the deformed lateral surfaces remain parallel to the original surfaces. The Neumann BC replacement is thus:

$$F_{\text{lateral tension}} = -\int_{\partial\Omega(\text{lateral surfaces})} T \cdot \eta \, dS.$$

(16)

*Modeling primary intention*

> Assumption 3: We assume that at static equilibrium, the various methods of closure (e.g. sutures, staples, cyanoacrylate, bandages) may only provide the traction $T$ in Eq. ( 8 ) as a vector directed in the $xy-$plane, i.e. parallel to the undeformed skin surface.

Because wound closure is a large deformation of the skin, we must transform surface normal vectors using Nanson's formula[8]:

$$\mathrm{da}\,\hat{n} = \mathrm{dA}\,J\,\boldsymbol{F}^{-T}\widehat{N},$$

( 17 )

where $F$ is the deformation gradient, $J$ is the determinant of $\boldsymbol{F}$, $\mathrm{da}$ is the deformed surface element area, $\hat{n}$ is the deformed surface normal, $\mathrm{dA}$ is the undeformed surface element area, and $\widehat{N}$ is the undeformed surface normal.

> Assumption 4: We assumed that after closure of the wound, the traction on the lateral faces of the wound are unchanged from what they are in normal healthy skin.

Therefore, we first calculated the traction equivalent to the imposition of Eqs. ( 12 ), ( 13 ), and ( 14 ) in normal healthy skin using Eq. ( 15 ). We then applied this traction to the wound geometry, and finally apply additional traction to close the wound. We reason that this approximation is valid when the tissue deficit is small relative to the body part, such that the pre- and post-closure skin tensions are similar. This condition is naturally satisfied when closure by primary intention is appropriate.

To estimate the wound closure traction, we applied a Dirichlet BC so that the wound edges meet in the middle. For a wound with reflection symmetry across the plane $x = a$, requiring that the exposed wound surfaces meet at $x = a$ ensures that the tension on the wound edges are equal and opposite:

$$DBC_{\text{primary intention}} = \{u_x = a\}.$$

( 18 )

We may then replace the Dirichlet BC with a traction:

$$F_{\text{primary intention}} = -\int_{\partial\Omega(\text{wound surface})} T \cdot \eta \, dS.$$

( 19 )

*Material depth contact model*

To model the contact of the wound edges after joining, we used a penalty-method contact model based on material depth[12]. The mathematical formulation is presented in the Supplemental Methods.

*Modeling primary intention in six stages*

We performed modeling for each geometry in six stages, enumerated here along with the corresponding variational form ($F_{\text{total}}$) and Dirichlet boundary conditions ($DBC$). Each stage depended on the result of the previous stage. For primary intention, we applied Dirichlet and Neumann BCs only on the middle third of the wound (y-axis) and only to a half its depth (z-axis).

1. Pre-tensioning:

$$F_{\text{total}} = F_{\text{constitutive}}$$

$$DBCs = DBC_{\text{lateral tension}} \text{ and } DBC_{\text{bottom}}.$$

2. Primary intention using a Dirichlet BC, switching from Dirichlet lateral BCs to Neumann lateral BCs:

$$F_{\text{total}} = F_{\text{constitutive}} + F_{\text{lateral tension}}$$

$$DBC = DBC_{\text{bottom}} \text{ and } DBC_{\text{primary intention}}.$$

3. Switching from the Dirichlet BC for primary intention to the Neumann BC:

$$F_{\text{total}} = F_{\text{constitutive}} + F_{\text{lateral tension}} + F_{\text{primary intention}}$$

$$DBC = DBC_{\text{bottom}}.$$

4. Adding the contact penalty, with $k_V = 1$ GPa:

$$F_{\text{total}} = F_{\text{constitutive}} + F_{\text{lateral tension}} + F_{\text{primary intention}} + F_{\text{contact, volumetric}}$$

$$DBC = DBC_{\text{bottom}}$$

5. Increasing $k_V$ to 5 GPa ($F_{\text{total}}$ and $DBC$ unchanged).

6. Increasing $k_V$ to 100 GPa ($F_{\text{total}}$ and $DBC$ unchanged).

*Characterizing fragility using the von Mises stress*

The von Mises stress $\sigma_{\text{von Mises}}$ is a metric of fragility[8], and correlates with the likelihood of the material breaking. It is defined by the equations:

$$\sigma_{\text{von Mises}} = \sqrt{\frac{3}{2} \boldsymbol{s}:\boldsymbol{s}}$$

$$\boldsymbol{s} = \boldsymbol{\sigma} - \frac{\text{tr}(\boldsymbol{\sigma})}{3}\boldsymbol{I}$$

$$\boldsymbol{\sigma} = J^{-1}\boldsymbol{P}\boldsymbol{F}^T.$$

( 20 )

To mitigate finite size effects, we calculated the 99[th] percentile of the von Mises stress in each mesh. In the sequel, we call this value the "maximal stress".

*Statistical and predictive modeling*

We used gradient-boosted decision trees[14] to predict the maximal stress based on the adjustable parameters. We binned the stresses into $q = 8$ quantiles in comparison to all simulations. We used a SoftMax loss function on the predicted category. One-vs-rest receiver operator characteristics[15] (ROCs) were plotted; macro-averaging is a simple mean of all quantiles for the sensitivity and specificity, whereas micro-averaging first sums the true positives, false positives, true negatives, and false negatives across quantiles, before computing these quantities.

*Software*

All code was written in Python 3.11, and simulations were performed on AMD EPYC 7763 processors in a Linux environment, while analysis was conducted on an M1 Max MacBook Pro. We used gmsh 4.11[5] for initial geometry generation, DOLFINx 0.7.2[2] and UFL 2021.1.0[9] for implementation of finite element models, NumPy 1.26.2[16] for numerical array processing, PyVista[17] for 3D visualization, matplotlib 3.8.2[18] for 2D plots, and XGBoost 2.0.3 for gradient-boosted decision trees.

**RESULTS**

We performed a total of 69,984 simulations for 11,664 geometries (six stages per geometry) with ten adjustable parameters (Table 1), to approximate a range of potential wound geometries and to test the robustness of our methods.

Out of the 69,984 simulations, 68,806 (98.3%) converged within 2 days of simulation on a single processor. In addition to traditional figures, we include interactive HTML 3D models of all converged simulations in our Supplemental Materials. We focused on wounds with reflection symmetry and did not model the wound ends. We limited primary intention to half the wound depth and the middle third of the wound long axis.

*Stress is concentrated at discontinuities in the material continuum, applied forces, and constraints*

We calculated the locations of maximum stress along paths through the tissue (Figure 2), and observed that stress is concentrated near discontinuities, either due to absence of tissue or due to a discontinuities in applied stress. Stress was concentrated along the wound long axis at the limits of where the primary intention boundary conditions were applied. Similarly, stress was concentrated near the thinnest portion of the skin along the wound short axis (x-axis). Finally, stress was concentrated in the stiff dermis and the most superficial portion of the epidermis.

*Approximating the function from simulation parameters to maximal stress*

We observed the following effects for the adjustable parameters on the stress. First, as a sanity check, 93.5% (10,819 out of 11,566; $p < 0.001$) of 'stiff' skin-type, curved wound simulations had greater stress than the 'soft' simulations, and the same held true for 79.9% (8,873 out of 11,105; $p < 0.001$) of wedge wound simulations. The remaining parameters are nonlinear and codependent, and this analysis is prone to confounding. Therefore we used gradient-boosted decision trees to fit predictive models of the maximal stress[14].

Using all the simulations where the maximal von Mises stress was greater than 0 (n = 68,592), we randomly split the data into training (n = 43,214), validation (n = 18,519), and test (n = 6,859) sets. Our goal here was to predict in which of 8 quantiles (i.e. 0-12.5% of all simulations, 12.5-25%, etc.) the maximal stress produced by a set of simulation parameters would fall. We adjusted the decision tree hyperparameters until the loss function plateaued (Figure 3A and B) and there was an acceptable amount of overfitting as measured by the training vs validation loss. We found that the wound depth was the most influential parameter in determining the maximal stress (Figure 3C), followed by the simulation stage. We evaluated the performance of our model on the test set with one-vs-rest ROCs (Figure 3D). The quantiles were well-balanced and therefore the micro- and macro-averaged ROCs coincide. The areas under the curve (AUCs) were above 0.85, indicating the decision tree was a sensitive and specific model for this classification task.

**DISCUSSION**

The simulations confirm intuition about primary intention wound mechanics. For missing tissue, stresses must conduct through the remaining tissue, causing increased deformation and concentration of stress near discontinuities (Figure 2). The stress was most heavily influenced by wound depth, whereas the other parameters we examined, such as wound width and dermal thickness were of relatively equal importance (Figure 3A). We note that using an adaptive Newton's method was essential to ensuring nonlinear solving convergence[6]. We created a predictive decision-tree to determine how much stress a given set of simulation parameters will cause (Figure 3B-D), yet such a model is limited to the range of available data.

Clinically, the quality of wound closure affects the aesthetics and medical efficacy of primary intention[19–21]. Given how ubiquitous primary intention is in the clinic, little has been done to simulate the mechanical aspects of wound closure. There is a large body of literature on measurement of intrinsic skin mechanical properties[22] as well as a wide variety of finite element models in use[23], however application of these methods to wound closure are limited[4,24]. Our hope is that by providing highly reproducible and effective methods, this work will help begin to fill in this gap in research.

To summarize our methods: first, we developed a method of replacing intrinsic skin tension originally imposed by Dirichlet BCs[7] with natural Neumann BCs, allowing us to model a wound in a skin patch in isolation while retaining the mechanical properties of the skin patch. We then justified a method of modeling primary intention which avoids costly contact mechanics using Dirichlet BCs, converting these into natural Neumann BCs. Finally, we adapted an existing method for contact mechanics[12] for use in DOLFINx.

The major limitation of this study is the use of preexisting material parameters. We did not perform a meta-analysis of the literature to obtain parameters for simulation, nor did we obtain *in vivo* data. It appears that skin material parameters are highly heterogeneous and method-dependent[22].

A potential goal would be to create a real-time, interactive algorithm which actively measures the mechanical properties of wounds during closure, and suggests optimal boundary conditions (i.e. suture placement, cyanoacrylate usage, bandage sizes and tensions) to minimize or equalize mechanical stresses in the skin. Real-time measurements would aid in precisely customizing wound closure to the patient, and combined with surgical robots could potentially enhance automation of wound closure.

**CODE AND DATA AVAILABILITY**

All code necessary to reproduce this work may be found in a GitHub repository (https://github.com/mikesha2/finite_element_skin). The code includes pre-built Docker containers with all necessary packages to execute the code, a Jupyter notebook (Main.ipynb) showing step-by-step implementation of the methods, another Jupyter notebook (Simulation Analysis.ipynb) reproducing all plots in this work, and a production-ready script (to_run.py) to reproduce any of the 68,804 simulations. Interactive 3D HTML plots files are available at Zenodo (doi: 10.5281/zenodo.10632859). Full simulation data are available upon request to the corresponding author (cms6712@psu.edu).

**ACKNOWLEDGEMENTS**

CMS is grateful to the Penn State Medical Scientist Training Program. Part of this work was performed while CMS was a guest at Argonne National Laboratory. CMS is also grateful for computational resources provided by Penn State Institute for Cyber Science and the Google Cloud research credits program.

**TABLES**

Table 1: Adjustable characteristics of models. We omitted models which exceeded 40,000 finite elements.

| Characteristic | Possible values | Description |
|---|---|---|
| wound shape | cylindrical, wedge | The wound shape in the plane of the skin was either a right circular prism (cylindrical) or a right isosceles triangular prism (wedge). |
| combined epidermal/dermal thickness | 0.5 mm, 1 mm, 2 mm, 4 mm | The total thickness of the epidermis and dermis, treated as a single material. |
| skin patch size factor | 1x, 2x, 3x | The amount of normal skin to the left and right of the wound (x-axis), as well as the length of wound and skin along the axis of the wound (y-axis) as a multiple of the epidermal/dermal thickness. |
| wound width factor | 0.3x, 0.5x, 0.8x | The width of the wound (x-axis) as a factor of the epidermal/dermal thickness. |
| subcutis thickness factor | 0.5x, 1x, 2x | The thickness of the subcutis as a factor of the epidermal/dermal thickness. |
| wound depth factor | 0.3x, 0.5x, 0.7x | The depth of the wound as a factor of the total skin thickness (epidermis + dermis + subcutis). |
| mesh size factor | 0.3x, 0.5x, 0.7x | A scaling parameter for the mesh sizes; larger values indicate finer meshes. |
| wound aspect factor (for cylindrical wounds) | 0x, 0.1x, 0.2x | For the cylindrical wound, a factor of the wound depth by which the cylinder is displaced in the superficial direction. |
| wound orientation (with respect to the material axes) | 0, 45º, 90º | The angle that the long axis of the wound makes with respect to the stiff material axis in the plane of the skin. |
| skin type (see Table 2) | soft, normal, stiff | The skin types that were defined by Flynn et al[7]. |

Table 2: Listing of coefficients used in the constitutive equations. The Mooney-Rivlin model was used for the subcutis, whereas the Fung models were used for the epidermis and dermis. Multiple sets of coefficients were tested for the Fung models to mimic normal, stiff, and soft skin; coefficient values are listed in this order. These values are reproduced from Tables 2 and 3 of Flynn et al[7].

| Model | Tissue | Coefficient | Values (kPa) |
|---|---|---|---|
| Mooney-Rivlin | Subcutis | $C_{10}$ | 0.4 |
| | | $C_{20}$ | 1.4 |
| | | $K$ | 50 |
| Fung | Dermis and epidermis | $c$ | 21.3, 42.6, 10.7 |
| | | $\mu_1$ | 17.8, 35.6, 8.9 |
| | | $\mu_2$ | 5.9, 11.8, 3.0 |
| | | $\mu_3$ | 5.9, 11.8, 3.0 |
| | | $\lambda_{11}$ | 1.0, 11.8, 0.5 |
| | | All other $\lambda_{ab}$ | 1.0, 2.0, 0.5 |
| | | $K$ | 250.0, 250.0, 250.0 |
| | | TSF | 1.10, 1.05, 1.15 |

**FIGURES**

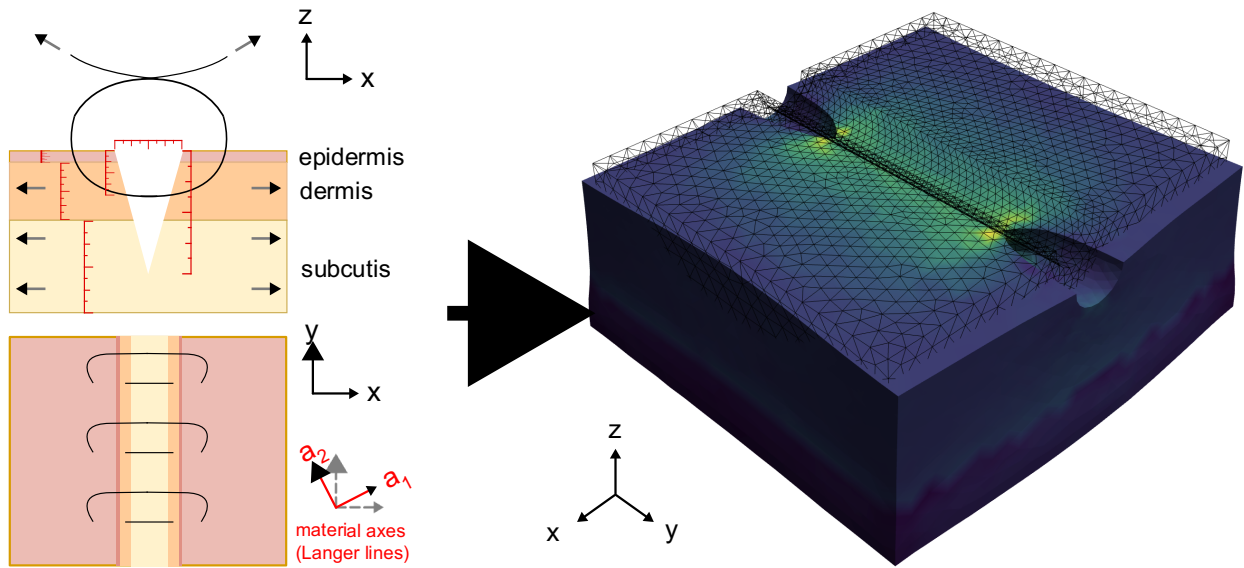

Figure 1: **Graphical abstract**. Simulation of soft tissues such as the skin is complicated by highly variable skin geometry, intrinsic tension, and nonlinear mechanical properties. The goal of this work is to introduce finite element and nonlinear regression methods for simulating primary intention wound closure, and to provide a guide for future simulations in the hopes of reducing the rate of wound dehiscence, tissue strangulation, and other complications as well as optimizing clinical procedures at the bedside.

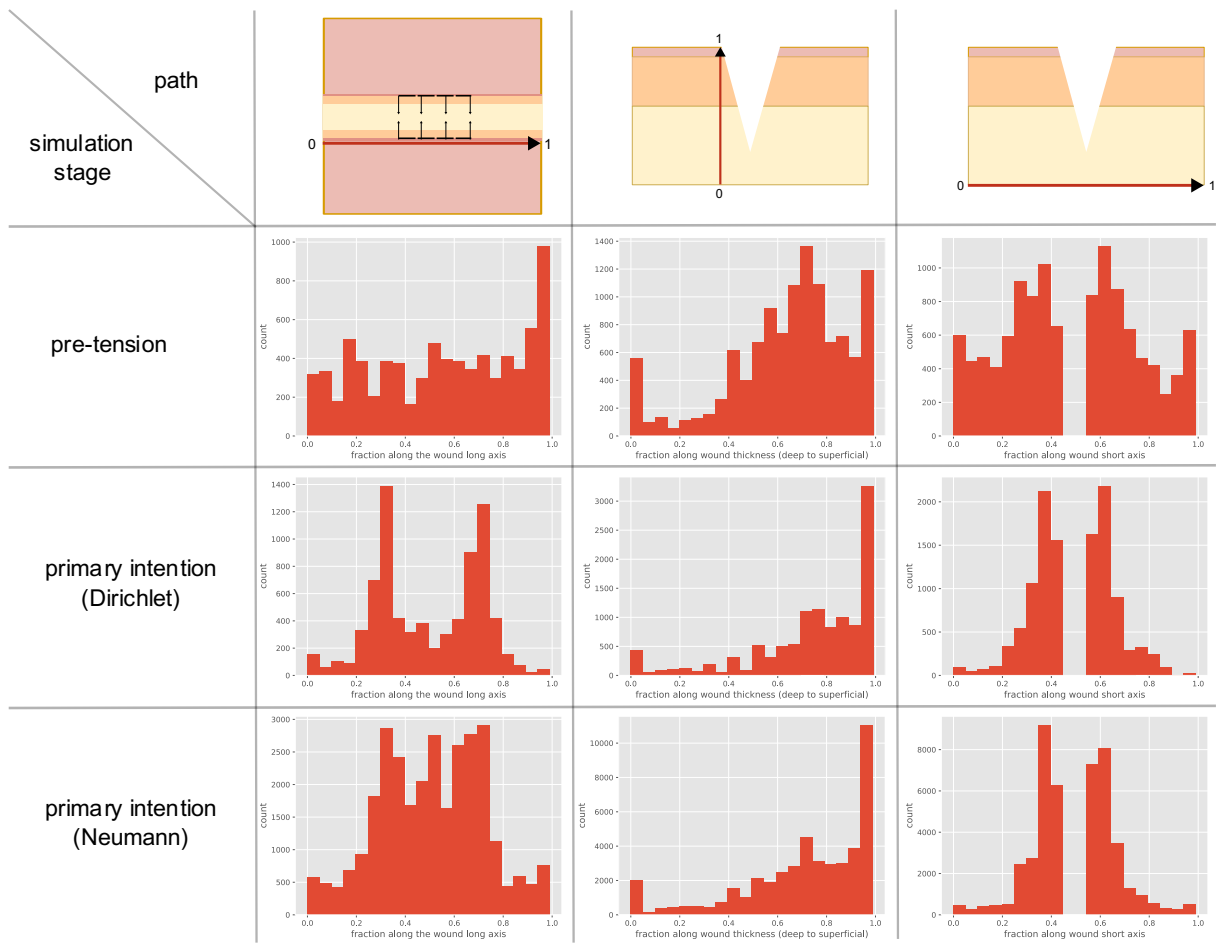

Figure 2: **Distribution of maximal stresses as a function of location within the wound.** For each axis and simulation stage, we calculated the distribution of locations of maximal stress across all simulations. The top row shows the path corresponding to the locations for which the maximum stress was calculated. The first column labels which simulation stage was considered (pre-tension: stage 1, Dirichlet primary intention: stage 2, Neumann primary intention: stages 3-6). For the first column of histograms, we observe that primary intention by Dirichlet BCs causes the stress to be concentrated near the limits of the primary intention, i.e. at 1/3 and 2/3 along the long axis of the wound. In contrast, Neumann BCs result in far lower

concentration of that stress. For the second column of histograms, the stress is concentrated in the superficial layers of the skin for all stages, as expected. For the final column of histograms, we observe that the presence of the wound causes that stress to be propagated all the way to the bottom of the tissue, and concentrated to each side of the wound. Even in the models in which primary intention BCs were not applied (pre-tension), there is increased stress at these boundaries.

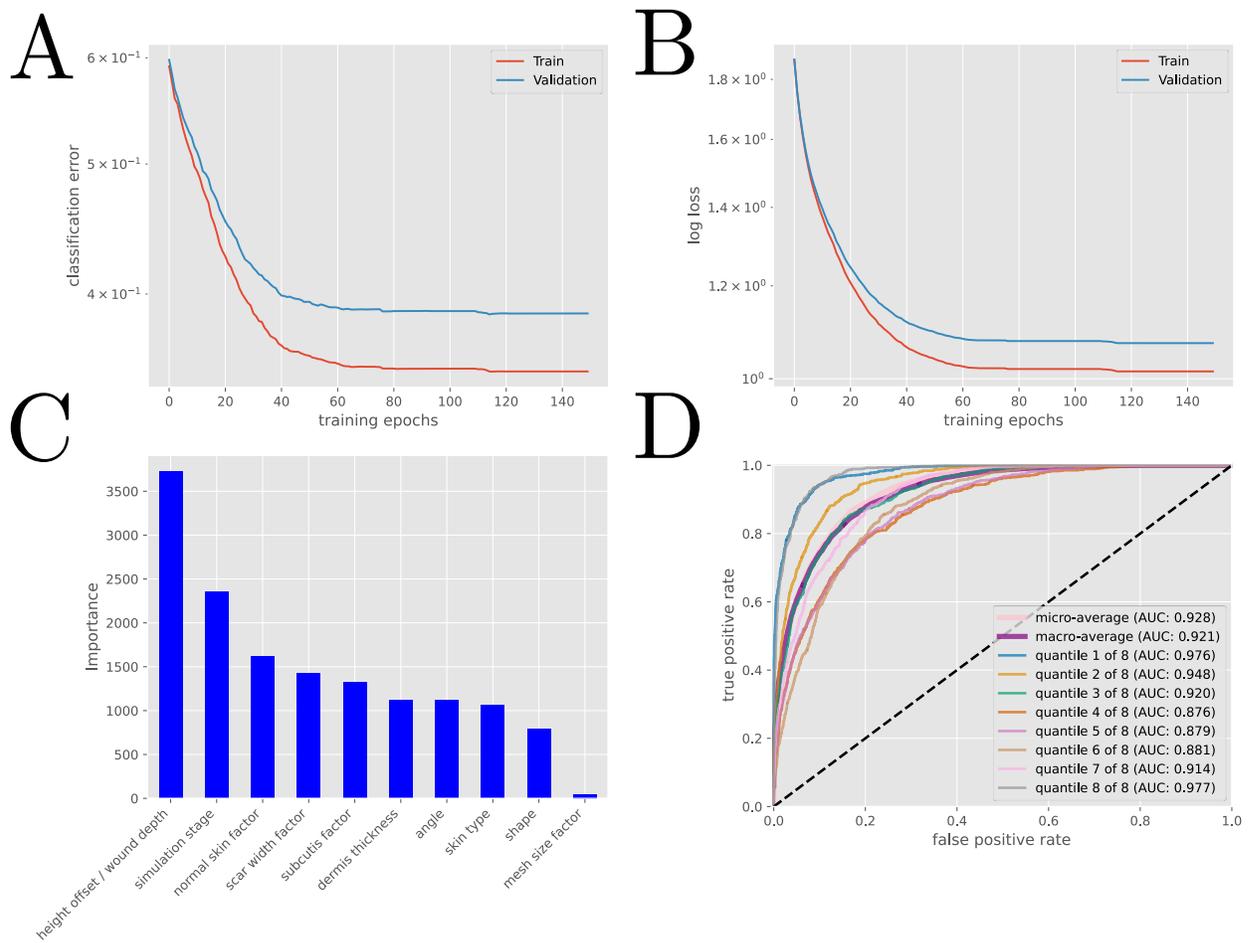

Figure 3: **A gradient-boosted decision tree predicts the maximal stress given simulation parameters.** (A) and (B) show the classification (SoftMax) error and log loss (cross-entropy) respectively as a function of training time, with both plateauing by epoch 100. Both metrics are similar between the training (n = 43,214) and validation (n = 18,519) sets, and therefore we concluded there was an acceptable degree of overfitting. (C) shows the estimated relative importance of each of the simulation parameters by a simple count of the number of times each given parameter appears as a decision in the tree. (D) shows the receiver operator characteristics (ROCs) for the binary classification of one quantile vs the rest on the test (n = 6,859) set. The micro- and macro-average ROCs are also plotted. Because of the high areas under the curve (AUCs), we conclude that the gradient-boosted decision tree is an appropriate choice for modeling the nonlinear dependence of the maximal von Mises stress as a function of the simulation parameters.